\documentclass[12pt]{iopart}

\usepackage{graphicx}
\newcommand{\bra}[1]{\langle #1 |}			
\newcommand{\ket}[1]{| #1 \rangle}
\newcommand{\proj}[2]{\ket{#1}\bra{#2}}

\newcommand{\EP}{(el)} 
\newcommand{\PP}{(pol)} 

\newcommand{\ham}{H}
\def\figHc{5cm}
\def\figWc{6.7cm}

\begin{document}
\title{ A density matrix approach to the dynamical properties of a two-site
Holstein model}
\author{S Paganelli and S Ciuchi} 
\address{Dipartimento di Fisica, Universit\`{a} dell'Aquila,
Via Vetoio,I-67100 L'Aquila, Italy}
\address{CRS SMC, INFM-CNR, Roma Italy}
\ead{\mailto{simone.paganelli@roma1.infn.it},\mailto{sergio.ciuchi@aquila.infn.it}}


\begin{abstract}
The two-site Holstein model represents a first non-trivial paradigm for 
the interaction between an itinerant charge with a quantum oscillator, a very
common topic in different ambits.
Exact results can be achieved both analytically and numerically, nevertheless
it can be useful to compare them with approximate, semi-classical techniques
in order to highlight the role of quantum effects.
In this paper we consider the adiabatic limit in which the oscillator
is very much slow than the electron. A density matrix approach is introduced for
studying the charge dynamics and the exact results are compared with 
two different approximations: a Born-Oppenheimer-based Static Approximation for the oscillator (SA) and a Quantum-classical (QC)
dynamics.
\end{abstract}

\maketitle
\section{Introduction}
\label{intro}

A common problem in chemical reaction and in solid state systems consists in tunneling charges between localized sites. The surrounding  crystal  or
molecular system interacts with the moving charge and can  hugely affect its
dynamics.
The competition between kinetic energy of the charge
and localization effects, due to local coupling with lattice vibration (phonons), produces the
small polaron, i.e. a charge dressed by a cloud of multiphonon processes, 
when the latter prevails \cite{tiablikov,holstein}. 

The  two-site cluster  is the minimal system in which competition 
of hopping between two sites and phonon localization effects takes place
and it
has been extensively studied since the pioneering work of Holstein
\cite{holstein}. Ground state properties
\cite{alexandrov,Feinberg90,firsov,cinesi} along with spectral properties
\cite{swain-cfracsol,alexandrov,demello,paganelciuk,ammberciu} and
correlation functions dynamics \cite{demello,herfortfilosofico} have been studied using both 
numerical and  analytical methods. Also for its simplicity, the two-site cluster has 
been
studied in more involved problems such as polaron formation in the 
presence of
double exchange \cite{ciuchi2} or in the presence of on-site electronic
repulsion \cite{acquarone}.
Recently, this system has also been adopted  to explain  \cite{FeinbergZazunovPRB06} 
conductance experiments in nanotubes \cite{leroynano}
The two-site system, introduced in the aforementioned mentioned cases 
of solid state physics, is relevant also in  the physics of organic 
materials \cite{SSH,organicbook,organicbook1,Sculten,DNA,twist}. 

In this paper, we  study the  electron and polaron reduced 
density matrix in order to characterize the transfer dynamics, its degree of coherence and how the temperature affects these dynamical properties.
In a previous work \cite{paganelciuk} we used an analytical technique, consisting in a mapping
into an effective anharmonic oscillator, for studying the spectral functions.
The same technique can be used also for getting exact results for the dynamics.  
Here, we focus on the adiabatic regime, in which the lattice or nuclei dynamics is much slower than that of the electron.
In this limit, several approximate techniques are usually 
adopted. Here we compare exact results with a Born-Oppenheimer static  approach (SA) and a Quantum-Classical (QC)  dynamics.
The comparison between different techniques can be done easily in the two-site 
system but it can be useful also for testing the validity of these methods to a more extended system.

The model is described in Sect. \ref{sec:model}. In Sect. \ref{sec:redmat}
we introduce the reduced density  matrix 
for the polaron and the electron. In Sect. \ref{sec:ana}
we describe the exact mapping, by means of Fulton-Gouterman transformations,
from the original electron-phonon problem into two single anharmonic oscillators.
Then the QC and SA approximation are described.
In \ref{sec:compar} some result are presented and the 
comparison between these three different techniques is discussed.
 
\section{The model}
\label{sec:model}

The model describes an electron,  in the tight
binding approximation, moving in a two-site lattice and interacting with
it by the local distortion of the lattice site.
The Hamiltonian is  \cite{firsov}
\begin{equation}\label{eqn:h2sitidispless}
    \ham=-J( c_1^\dag c_2+c_2^\dag c_1)+\omega_0(a^\dag_1a_1+a^\dag_2 a_2)-g[c^\dag_1c_1
    (a^\dag_1+a_1)+c^\dag_2c_2(a^\dag_2+a_2)]
\end{equation}
$c_j^\dag$ and $a_j^\dag$ are, respectively,  the electron and phonon creation
operators.
The strength of the
electron-phonon interaction is given by the constant $g$ and $J$ is the
electron hopping integral. Here we  consider 
one local  oscillator per site with frequency $\omega_0$.

We can reduce the the degrees of freedom introducing the coordinates
corresponding to the center of mass and the relative displacement. 
The center of mass Hamiltonian consists in a displaced
oscillator and can be separated from the part
depending on the relative coordinate ($a = (a_1-a_2)/\sqrt{2}$).
In the following discussion we shall limit ourselves to study only the latter   
\begin{equation}\label{eqn:holstein}
  H = \omega_0 a^\dag a-J \sigma_x-\tilde{g}\sigma_z(a^\dag+a).
\end{equation}

In (\ref{eqn:holstein}) $\tilde{g}=g / \sqrt{2}$ and a pseudo-spin notation 
has been used by introducing the Pauli matrices $\sigma_z = c_1^\dag c_1 - c_2^\dag c_2 $
and $\sigma_x = c_1^\dag c_2 + c_2^\dag c_1$.

The Hamiltonian (\ref{eqn:holstein}) has a very general form and, 
even if is has been derived for a two-site cluster, 
it is suitable to describe a very wide class of problems
(for example a two level system interacting with a single
optical mode \cite{swain-cfracsol}).

Beside the temperature, we can choose two parameters that characterize the
model  i) the bare e-ph coupling constant $\lambda=g^2/(\omega_0 J)$ given by
the ratio of the  polaron energy ($E_p=-g^2/\omega_0$) to the hopping
$J$ and ii) the adiabatic ratio $\gamma = \omega_0/J$. 
In terms of these parameters we can define a weak-coupling $\lambda<1$ and
strong coupling $\lambda>1$ regimes as well as an adiabatic $\gamma<1$ or
anti-adiabatic $\gamma>1$ regimes.
Notice that,  in the so called atomic ($J=0$) limit, the coupling's strength is better
described by another constant  i.e. $\alpha=\sqrt{\lambda/(2 \gamma)}$ 

In the atomic limit the Hamiltonian is diagonalized by the
so-called Lang-Firsov (LF) transformation \cite{lang} 
\begin{equation}\label{eqn:lfsep}
    D=e^{\alpha \sigma_z (a^\dag-a)}.
\end{equation}
This transformation shifts the phonon operators by a quantity $\alpha$, while
the electron operator is transformed into  a new fermionic one associated to
a quasi-particle, called polaron, with energy $E_p$. 
It can be shown that $\alpha^2$ is the mean number of phonons in the polaron cloud.

By applying the LF transformation $\bar{H}_{0}=D^\dag H_0 D$,
the atomic Hamiltonian 
$H_0 = \omega_0 a^\dag a -\tilde{g}\sigma_z (a^\dag+a)$ becomes
\begin{equation}\label{eqn:AtomicLimit}
    \bar{H}_{0}=\omega_0 a^\dag a+E_p/2,
\end{equation}
the eigenvalues $E_n=\omega_0 n+E_p/2$ correspond to the two-fold degenerate
eigenvectors $ \ket{\psi_n^j,j}= D \ket{n,j}=\bar{c}_j^\dag \ket{n}$ were the
index $n=0,\ldots,\infty$ refers to the photon number, $j=1,2$ to the electron
site number and $\bar{c}_j^\dag$ is the polaron creation operator $\bar{c}_j^\dag=D
c_j^\dag D^\dag=c_j^\dag exp\{(-1)^j \alpha(a^\dag-a)\}$.

In the case of finite $J$, the hopping term is not diagonalized by
(\ref{eqn:lfsep}) and  the new Hamiltonian $\bar{H}=D^\dag H D$
becomes
\begin{equation}\label{eqn:parappa}
    \bar{H}=\omega_0 a^\dag a -J(\sigma_x \cosh ({2 \alpha 
    (a^\dag-a)}) +i\sigma_y \sinh ({2 \alpha (a^\dag-a)}) )+E_p/2.
\end{equation}

Depending on the choice of the parameters, the problem could be better described
by a electron or polaron excitation picture. In particular, in the weak
coupling limit, both the small polaron and the electron are good quasiparticles
while, in the intermediate and strong coupling regimes, the polaron behaviour
prevails \cite{robin,paganelciuk}. 

\section{Reduced Density matrix}
\label{sec:redmat}

Hereafter, we shall assume that charge and oscillator are initially separate, being the former localized on the first site and the latter in a mixed thermal state.
The corresponding density matrix is
\begin{equation}
\rho(0)=\sum_n \frac{e^{-\beta \omega_0 n}}{Z}
\proj{\phi_n}{\phi_n}\otimes\proj{1}{1},
\end{equation}
where we used the notation $\ket{1}=c_1^\dagger \ket{0}$. The state
$\ket{\phi_n}$  depends on the choice of the initial preparation \cite{lucke}, in this paper we  study two different situations obtained from two
different limiting regimes:

\begin{enumerate}
\item \emph{electronic preparation} \EP: The first case corresponds to an initial free Hamiltonian ($g=0$) where
the oscillator is at its thermal equilibrium. In this case $\ket{\phi_n}=\ket{n}$.
\item \emph{polaronic preparation} \PP: In the second case we start from the atomic limit ($J=0$). Here the
oscillator is displaced  and the basis in
which it thermalizes is $\ket{\phi_n}=\ket{\psi^1_n}$  i.e. the displaced
oscillator eigenbasis.
\end{enumerate}

The dynamics is obtained by switching on  $g$, in the first case, and $J$, in the second one, and letting evolve the density matrix with 
the Hamiltonian   (\ref{eqn:holstein}) $\rho(t)=e^{-i Ht} \rho(0) e^{i H t}$.

Tracing over the oscillator degree of freedom, we obtain the electron reduced density matrix
\begin{equation}
\label{rhoel} 
    \rho^{(el)}(t)=\Tr_{ph}\{\rho(t)\},
\end{equation}
which, in terms of the oscillator's number states, is
\begin{equation} 
\label{rhoel1}
\rho^{(el)}(t) =\sum_{n,m} \frac{e^{-\beta \omega_0 n}}{Z}
    \bra{m}e^{-i  {H} t}\proj{n,1}{n,1}e^{i
    {H}t}\ket{m},
\end{equation}

To  characterize the motion of the polaron we cannot  reduce the
density matrix  by  tracing out the phonon degrees of
freedom, this is because the polaron itself contains phonons.
In order to understand better the polaron dynamics, let us first apply a Lang-Firsov transformation, the new fermionic particle corresponds to a polaron, so the 
density matrix  with the initial localized polaron can be written as
\begin{equation}\label{eqn:poldensm}
\label{rhopol} 
\rho^{(pol)}(t) = Tr_{ph}\{D_R^\dag\rho(t)D_R \},
\end{equation}
and reads in terms of the oscillator's number state as
\begin{equation}
\label{rhopol1}
\rho^{(pol)}(t) =\sum_{n,m} \frac{e^{-\beta \omega_0 n}}{Z}
    \bra{m}e^{-i \bar{H} t}\proj{n,1}{n,1}e^{i
   \bar{H}t}\ket{m}.
\end{equation}

The diagonal elements of the reduced density matrix, in the site basis,
represent the population of each site while the off-diagonal elements are called
\emph{coherences} and represent the quantum interference  between localized
amplitudes. An important quantity, which is also base-independent, is the
so-called \emph{purity} defined as
\begin{equation}
\label{purity}
P=\Tr\{\tilde{\rho}^2\} .
\end{equation}
It is easy to see that $1/2 \leq P \leq 1$  and it measures how much the state is pure ($P=1$ if and only if the state is pure and
$P=1/2$ when it is maximally mixed).


\section{Analytical approaches}\label{sec:ana}

\subsection{The adiabatic, static, approximation}\label{sub:adiab}
The case, in which a light quantum particle interacts with  much more massive particles,  is
very common in solid state and molecular physics.
We discuss the \emph{adiabatic regime}, meaning that in a characteristic time for the  the light 
particle dynamics the heavy degrees of freedom can be considered approximately quiet. Here we describe the SA approach in its basic formulation for the dynamics.

The Hamiltonian (\ref{eqn:holstein}) can be written 
in the coordinate-momentum representation 
\begin{equation} \label{eqn:Hadiab}
    \ham=\frac{p^2}{2 m}+\frac{m
\omega_0^2}{2}x^2-\frac{\bar{g}}{\sqrt{2}}x\sigma_z-J\sigma_x-\omega_0,
\end{equation}
with $\bar{g}=g \sqrt{2m \omega_0}$.
In the adiabatic limit ($\gamma\ll1$), the phonon is much slower than the electron
(heavy phonon and large electron tunnelling amplitude) and one can neglect the phonon kinetic
term in (\ref{eqn:Hadiab}). This is the well known Born-Oppenheimer approximation. In practice, 
 it consists in studying  the electronic problem with $x$ as a classical parameter.

Within this approximation, we put $\omega_0=\sqrt{k/m}\rightarrow 0$ ($m$ is the ion mass) and  the Hamiltonian becomes
\begin{equation}\label{eqn:relham}
    \ham_{ad}=\frac{k}{2}x^2-\frac{\bar{g}}{\sqrt{2}}x \sigma_z-J\sigma_x.
\end{equation}
The eigenvalues can be expressed trough the classical displacement $x$
\begin{equation}
\label{eqn:adiabatic_energies}
V_\pm(x)=\frac{k}{2}x^2 \pm \Omega(x),
\end{equation}
with $\Omega(x)=\sqrt{\frac{\bar{g}^2}{2}x^2+J^2}$.
The lowest branch ($-$) of  (\ref{eqn:adiabatic_energies})
defines an adiabatic potential which has a minimum at $x=0$ as far as
$\lambda<1$ while for $\lambda>1$, it becomes double well potential with minima at 
$x=\pm x_m=\pm \sqrt{\frac{\bar{g}^2}{2 k^2}-\frac{2 J^2}{\bar{g}^2}}$, in this case  the electron is mostly localized on a
given site. The  quantum fluctuations are able to restore the symmetry
in analogy to what happens for an infinite lattice \cite{Lowen}. It is worth 
noticing that, in this limit, Hamiltonian  (\ref{eqn:holstein}) is equivalent to
the adiabatic version of the spin-boson Hamiltonian \cite{weiss,petruccione}.

The temporal evolution  is given by
\begin{equation}\label{eqn:evoluzio}
    e^{-i \ham_{ad} t}=e^{-i \frac{k x^2}{2} t}[\cos \Omega(x) t +i
    (\frac{\bar{g}x}{\sqrt{2}\Omega(x)}\sigma_z+\frac{J}{\Omega(x)}\sigma_x)\sin 
\Omega(x) t],
\end{equation}
so the density matrix dynamics can be explicitly calculated

The electronic initial preparation, corresponds to the density matrix 
\begin{equation}
\rho(0)=\proj{1}{1}\sqrt{\frac{k\beta }{2 \pi}} \int dx \,
e^{-\frac{\beta k}{2} x^2}\proj{x}{x},
\end{equation}
tracing out the phonon we obtain the electron reduced density matrix with elements
\begin{eqnarray}
\rho^{(el)}_{2,2}&=&\sqrt{\frac{\beta J \lambda}{2 \pi}}
\int du \, e^{-\frac{\beta J \lambda}{2} u^2}
\frac{\sin^2(J t \sqrt{u^2 \lambda^2+1)}}{1+\lambda^2 u^2}\nonumber\\
\rho^{(el)}_{1,2}&=&-i\sqrt{\frac{\beta J \lambda}{2 \pi}}
\int du \, e^{-\frac{\beta J \lambda}{2} u^2}
\frac{ \sin(2 J t \sqrt{u^2 \lambda^2+1}) }{2\sqrt{\lambda^2 u^2+1}}
\end{eqnarray}
where the scaled lenght $u=x k \sqrt{2}/\bar{g}$ was introduced.

In the same way we can introduce the polaronic preparation
\begin{equation}
\rho(0)=\proj{1}{1}e^{-\frac{\beta \bar{g}^2}{4k}    }\sqrt{\frac{k\beta }{2 \pi}} \int dx
e^{-\beta(
    \frac{k}{2}x^2-\frac{\bar{g}}{\sqrt{2}}x)} \proj{x}{x},
\end{equation}
It is worth stressing that, in the adiabatic limit, we cannot define the polaronic dynamics, as 
introduced in (\ref{eqn:poldensm}), because the operator $D$ is not defined for $\omega_0=0$, so
we better have to talk about an electronic dynamics with an initial polaronic preparation.
The corresponding reduced density matrix is

\begin{eqnarray}
\rho^{(pol)}_{2,2}&=&\sqrt{\frac{\beta J \lambda}{2\pi}}
    \int_{-\infty}^\infty du\,e^{-\frac{\beta J
    \lambda}{2}(u-1)^2}
  \frac{\sin^2(J  t \sqrt{(u \lambda)^2+1} ) }{ (u \lambda)^2+1}
\nonumber\\
\rho^{(pol)}_{1,2}&=&\sqrt{\frac{\beta J \lambda}{2\pi}}
    \int_{-\infty}^\infty du\, e^{-\frac{\beta J\lambda}{2}(u-1)^2}
     \left[ \frac{u \lambda\sin^2 (J t \sqrt{(\lambda u)^2+1}) }{((\lambda u)^2+1)}
       \right.\nonumber\\
    &-&\left.i \frac{\sin(2 J t \sqrt{(\lambda u)^2+1})}
    {2 \sqrt{(\lambda u)^2+1}}  \right]
\end{eqnarray}

It is possible to demonstrate that $\rho^{(pol)}_{2,2}$  is actually the adiabatic limit   
of the diagonal element of the reduced  polaronic density matrix, while this is not true for the 
off-diagonal elements.

We want stress that, in the SA approach, the phonon is completely \emph{static} because its momentum $p$ has been  neglected. Here, only the
initial phonon distribution plays a role, but during electron hopping, oscillator is taken to be fixed.

\subsection{A quantum-classical dynamics approximation}
\label{app:QC}
In the adiabatic limit, the slow variables can be considered as classical and a mixed quantum-classical dynamics can be introduced.
In the past, a lot of schemes for quantum-classical dynamics has been proposed, for example starting from the Born-Oppenheimer (SA) adiabatic approximation for the ground state at each step and using a density functional 
Hamiltonian \cite{car-parrinello,selloni}.
Another approach, good for a  short time dynamics, consists in a mapping from the Heisenberg equations
to a classical evolution by an average over the initial condition\cite{golosov,stock:1561}.
Some schemes are based on the evolution of the density matrix coupled to 
a classical bath \cite{berendsen,morozov}. A systematic expansion over the mass ratio has also 
been done, starting from partial Wigner transform of the Liouville operator, in  \cite{ciccotti1999,ciccotti2001JChPh,sergi2006JChPh}.
The QC approximation we use is essentially that of refs. \cite{berendsen,morozov}.

Let us now consider Hamiltonian (\ref{eqn:Hadiab}), where $x$ and $p$  are now two classical variables  which can be represented as the components of  a vector
$\mathbf{u}$. Then a QC vector can be introduced  
\begin{equation}
\mathbf{v}=\mathbf{u}\otimes \mathbf{\sigma}=\left(
\begin{array}{c}
x\\
p\\
\sigma_x\\
\sigma_y\\
\sigma_z
\end{array}
\right).
\end{equation}
The classical variables evolve with the Ehrenfest equations, defined for
the averages quantities  
\begin{equation}
\left\{
\begin{array}{ccl}
\dot{x}&=&\frac{p}{m}\\
\dot{p}&=&\frac{m\omega_0^2}{2}x-\frac{\bar{g}}{\sqrt{2}}\left\langle \sigma_z\right\rangle ,
\end{array}
\right.
\end{equation}
while the quantum variables evolves in the Heisenberg picture 
\begin{equation}
\left\{
\begin{array}{ccl}
\dot{\sigma}_x &=& -\sqrt{2} \bar{g} x \sigma_y        			\\
\dot{\sigma}_y &=&  \sqrt{2} \bar{g} x \sigma_x  -2J \sigma_z	\\ 
\dot{\sigma}_z &=& 	                              2J \sigma_y
\end{array}
\right..
\end{equation}
To give a unified description of the overall evolution, we define a Liouvillian 
operator $\mathcal{L}=\mathcal{L}_x+\mathcal{L}_p+\mathcal{L}_\sigma$
with 
\begin{equation}
\mathcal{L}_\sigma=-i \left(
\begin{array}{ccc}
 0 									& -\sqrt{2} \bar{g} x  	& 0 						\\
\sqrt{2} \bar{g} x	& 		0									& -2J 					\\
0										& 			2J 							&   0  
\end{array}
\right)
\end{equation}
and $\mathcal{L}_x = \dot{x} \frac{\partial}{\partial x} $ 
$\mathcal{L}_p = \dot{p} \frac{\partial}{\partial p}$.
So, the time evolution is given by  
\begin{equation}
\mathbf{v}(t)=e^{i \mathcal{L}t}\mathbf{v}(0).
\end{equation}

The numerical integration can be implemented using the symetrized Trotter breakup formula \cite{wan:4447,deraedt-trotter}
\begin{equation}
\mathbf{v}(t)\simeq \left(
e^{i \mathcal{L}_\sigma\frac{\epsilon}{2}}
e^{i \mathcal{L}_p     \frac{\epsilon}{2}}
e^{i \mathcal{L}_x           \epsilon}
e^{i \mathcal{L}_p     \frac{\epsilon}{2}}
e^{i \mathcal{L}_\sigma\frac{\epsilon}{2}}
\right)^N \mathbf{v}(0)
\end{equation}
with $\epsilon=t/N$. 
All the density matrix elements, can be expressed in terms of elements of $\mathbf{v}(t)$.

This approach, in contrast with SA, takes into account  the dynamics of the phonon, even if it remains classical.

\subsection{Exact diagonalization}
\label{analytics}

As shown by Fulton and Gouterman \cite{FG}, a two-level system coupled to an
oscillatory system in such a manner that the total Hamiltonian displays a
reflection symmetry, may be subjected to a unitary transformation which
diagonalizes the system with respect to the two-level subsystem
\cite{wagner,wagner3,FG}.
This method can be generalized to the N-site situation, if the symmetry of
the system is governed by an Abelian group \cite{wagner3}.

In particular, an analytic method for calculating the Green functions of the two-site Holstein model is
given in \cite{ciuchi2,paganelciuk}. Here the Hamiltonian is diagonalized in the fermion
subspace by applying a Fulton Gouterman (FG) transformation. So the initial problem is  mapped into
an effective  anharmonic oscillator model. It is possible to introduce  different FG transformations
for the electron and the polaron. The resulting problem results to be very simplified and 
very  suitable to be numerically implemented. Analytical continued-fraction results, exist
for the electron case\cite{paganelciuk,ciuchi2}.

In this section, we briefly recall the FG transformations method. 
The density matrix elements are given explicitly in terms of effective Hamiltonians and calculated
by means of exact diagonalization

The FG transformation  we use for the electronic case is 
\begin{equation}\label{eqn:can}
  V = \frac{1}{\sqrt{2}}\left(%
\begin{array}{cc}
  1  & (-1)^{a^\dag a}\\
  -1 & (-1)^{a^\dag
a}\\
\end{array}
\right),
\end{equation}
the new Hamiltonian $\tilde{H}=V H V^{-1}$ becomes diagonal in the 
electron subspace
\begin{equation}\label{eqn:acca}
    \tilde{H}=\left(
    \begin{array}{cc}
  H_+  & 0\\
  0 & H_- \\
\end{array}\right)
\end{equation}
the diagonal elements, corresponding  to the bonding and antibonding 
sectors of the electron subspace, being two purely phononic Hamiltonians 
\begin{equation}\label{eqn:hamph}
    H_\pm=\omega_0 a^\dag a \mp J(-1)^{a^\dag a}-\tilde{g}(a^\dag+a).
\end{equation}
The operator $(-1)^{a^\dag a}$ is the reflection operator in the vibrational 
subspace and it satisfies the condition
$(-1)^{a^\dag a}a(-1)^{a^\dag a}=-a$.
A wide study of the eigenvalue problem was carried out
in \cite{herfort1} both numerically and analytically by a variational method, 
extending the former results given in \cite{ranninger}.
In \cite{herfort1} $  H_\pm $ is approximately diagonalized by applying 
a displacement, the dynamics is reconstructed by 
the calculated eigenvectors and energies. 

The evaluation of the polaron Green function can be done on the same footings
but the expression involves also the non diagonal elements of the resolvent
operators causing  an exponential increasing of the numerical calculations.

To avoid this problem, we first perform  the FG transformation   
and then apply, on the resulting Hamiltonian (\ref{eqn:parappa}),
a different FG transformation 
\begin{eqnarray}\label{eqn:can1}
  V_1 = \frac{1}{\sqrt{2}}\left(
\begin{array}{cc}
  1             & -(-1)^{a^\dag a}\\
  (-1)^{a^\dag a} &   1\\
\end{array}
\right).
\end{eqnarray}
The new Hamiltonian $\tilde{H}_{LF}=V_1 \bar{H} V_1^{-1}$ is
\begin{equation}\label{eqn:acca1}
    \bar{H}_{LF}=\left(
    \begin{array}{cc}
  \bar{H}_+  & 0\\
  0 & \bar{H}_- \\
\end{array}\right)
\end{equation}
where
\begin{equation}\label{eqn:hbarra}
\bar{H}_\pm =\omega_0 a^\dag a+J(-1)^{a^\dag a} e^{\mp 2\alpha (a^\dag-a)}+E_p/2
\end{equation}
is real and symmetric but not tridiagonal in the basis of the harmonic
oscillator, the matrix elements of $\bar{H}_\pm$ are given in \cite{paganelciuk}. 

In order to write down the density matrix elements, let us introduce the following notation:
\begin{eqnarray}\label{eqn:risolventiel}
    R_{m,n}^{(\pm)}(t) &=&\bra{m}e^{-i\ham_\pm t}\ket{n}\\
    \bar{R}_{m,n}^{(\pm)}(t)&=&\bra{m}e^{-i\bar{\ham}_\pm t}\ket{n},
\end{eqnarray}
\begin{eqnarray}
N_{1,1}^{m,n}(t)&=&\bra{m,1}e^{-i\ham t}\ket{n,1}=
    \frac{1}{2}\left[ R_{m,n}^{(+)}(t)+R_{m,n}^{(-)}(t)\right]\nonumber\\
&=&(-1)^{n+m}N_{2,2}^{m,n}(t) \\
N_{2,1}^{m,n}(t)&=&\bra{m,2}e^{-i\ham t}\ket{n,1}=
    \frac{(-1)^m}{2}\left[ R_{m,n}^{(+)}(t)-R_{m,n}^{(-)}(t)\right]\nonumber\\
&=&(-1)^{n+m}N_{1,2}^{m,n}(t),
\end{eqnarray}
\begin{eqnarray}
M_{1,1}^{m,n}(t)&=&\bra{\psi_m^{1},1}e^{-i\ham t}\ket{\psi_n^{1},1}=
=\frac{1}{2}\left[\bar{R}_{m,n}^{(+)}(t)+(-1)^{m+n} \bar{R}_{m,n}^{(-)}(t)\right] \\
M_{2,1}^{m,n}(t)&=&\bra{\psi_m^{2},2}e^{-i\ham t}\ket{\psi_n^{1},1}=\frac{1}{2}\left[(-1)^{n}\bar{R}_{m,n}^{(-)}(t)-(-1)^{m} \bar{R}_{m,n}^{(+)}(t)\right].
\end{eqnarray}

The reduced electron density matrix elements are
\begin{eqnarray}
    \rho^{(el)}_{1,1}(t)&=& \sum_{n,m} \frac{e^{-\beta \omega_0 n}}{Z} |N_{1,1}^{m,n}(t)|^2\nonumber 	\\
	\rho^{(el)}_{2,1}(t)&=&\sum_{n,m} \frac{e^{-\beta \omega_0 n}}{Z}
     N_{2,1}^{m,n}(t) N^{ * m,n}_{1,1}(t), 
\end{eqnarray}
the calculation for the polaron  case gives
\begin{eqnarray}
    \rho^{(pol)}_{1,1}(t)&=&
    \sum_{n,m} \frac{e^{-\beta \omega_0 n}}{Z}
    |M_{1,1}^{m,n}(t)|^2 \nonumber \\
    \rho^{(pol)}_{2,1}(t)&=&
    \sum_{n,m} \frac{e^{-\beta \omega_0 n}}{Z}
     M_{2,1}^{m,n}(t) M^{ * m,n}_{1,1}(t).
\end{eqnarray}

\section{Comparison between different techniques}\label{sec:compar}

\begin{figure}[htbp]
\begin{center}
\includegraphics[width=\figWc,height=\figHc,angle=0]{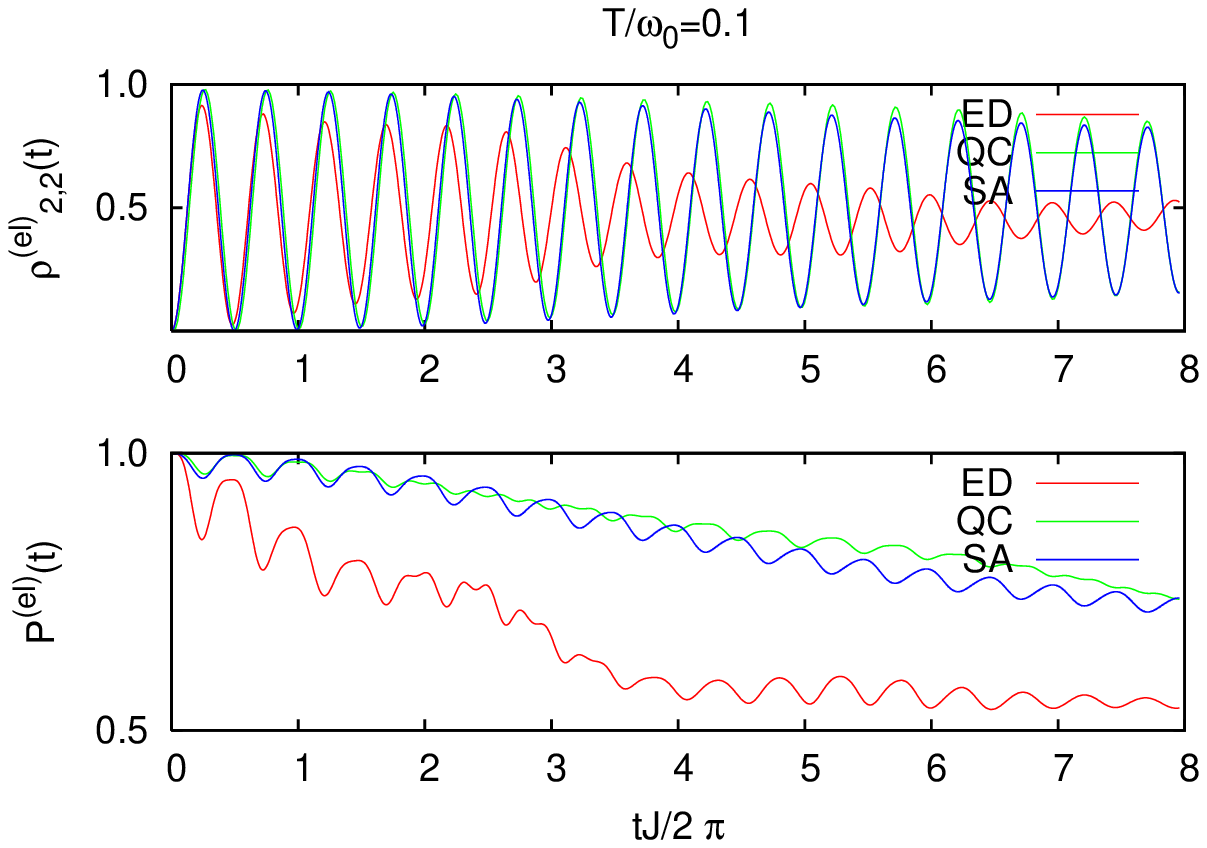}
\includegraphics[width=\figWc,height=\figHc,angle=0]{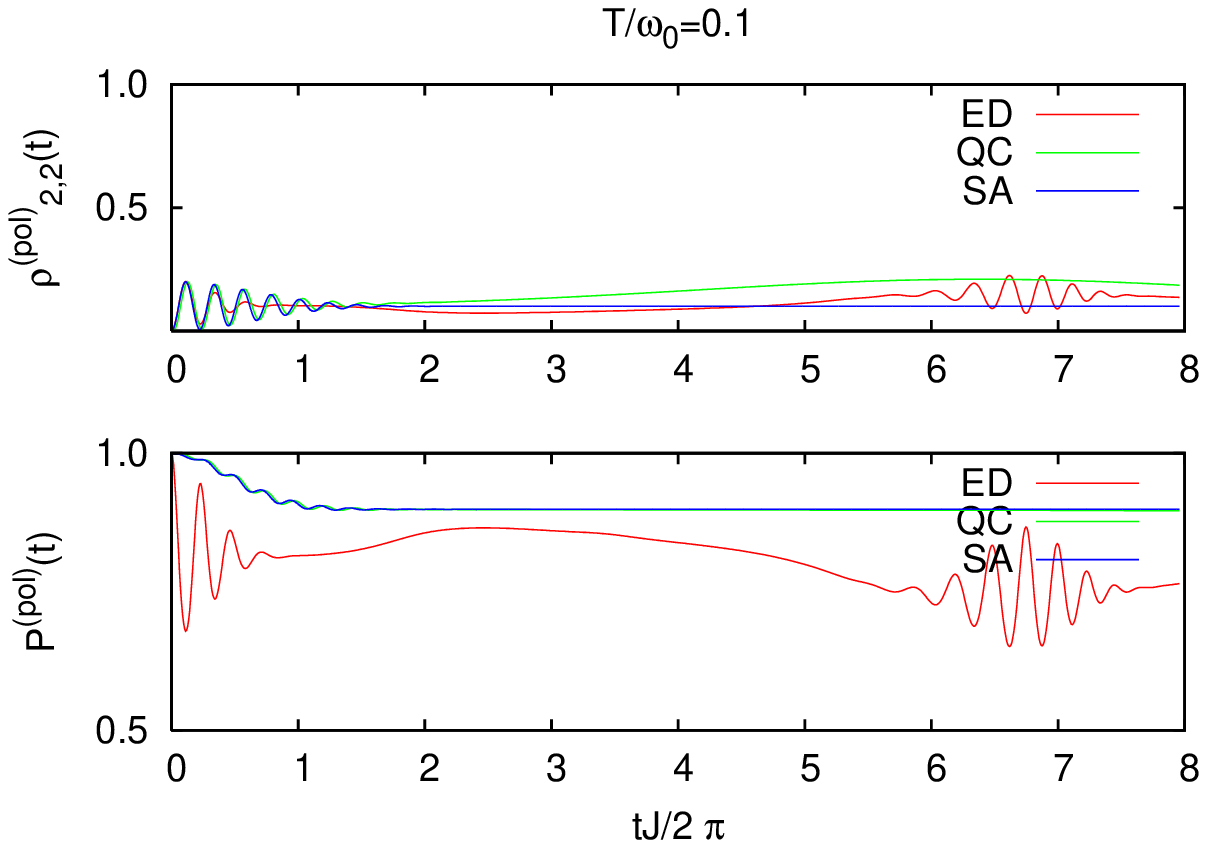}
\includegraphics[width=\figWc,height=\figHc,angle=0]{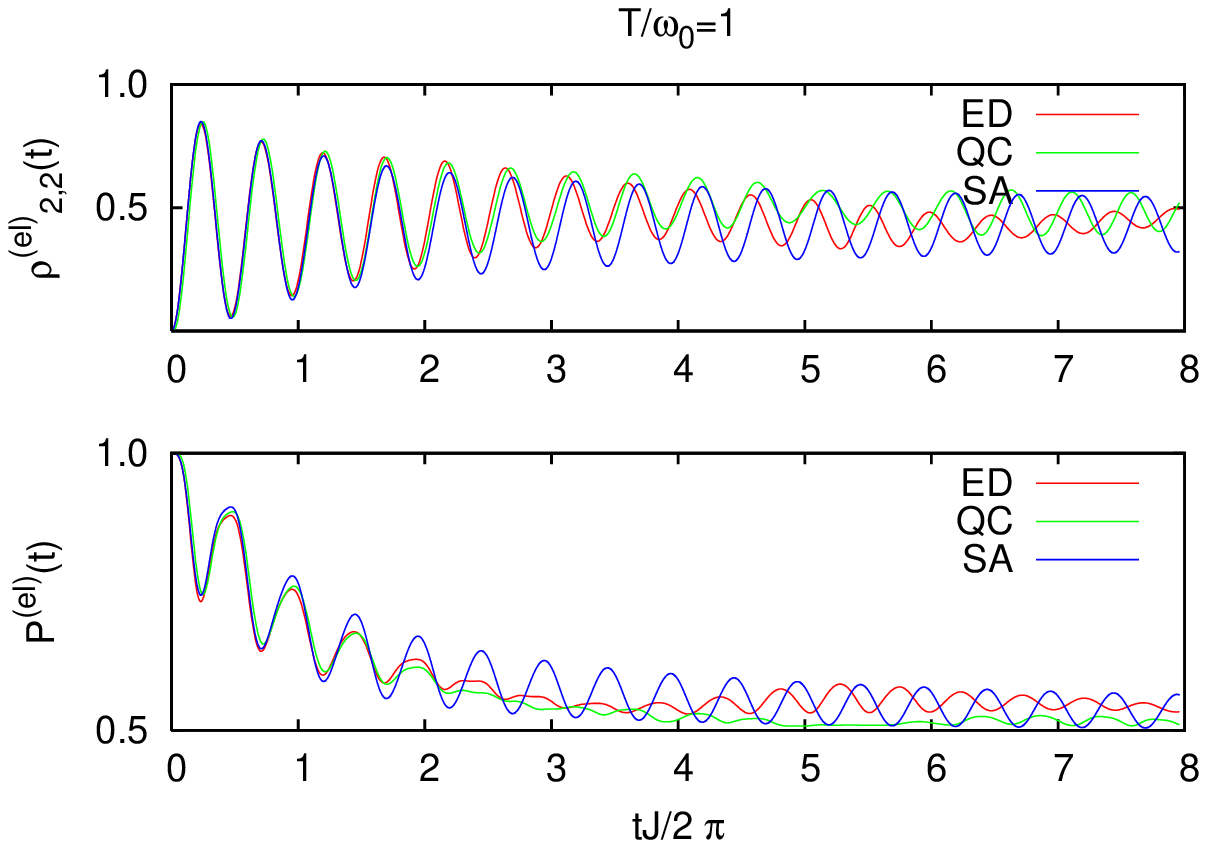}
\includegraphics[width=\figWc,height=\figHc,angle=0]{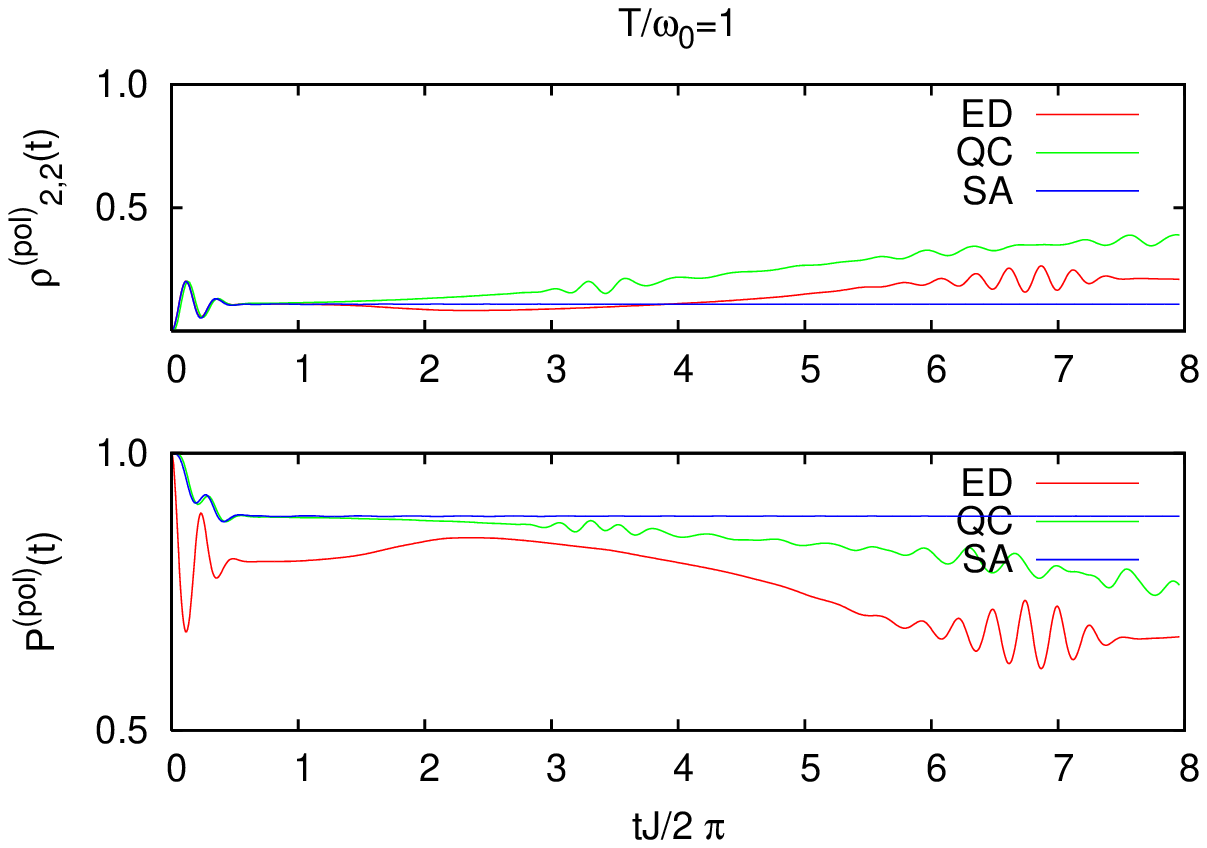}
\includegraphics[width=\figWc,height=\figHc,angle=0]{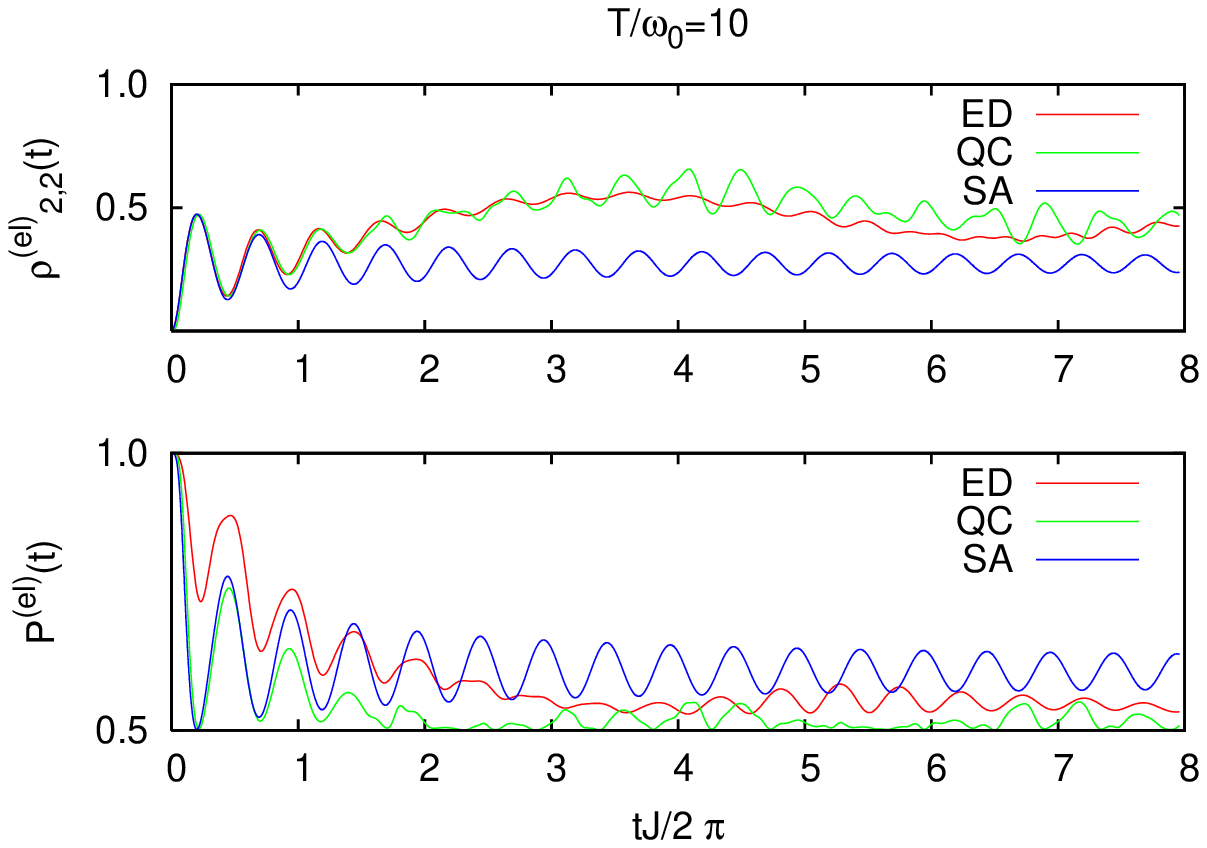}
\includegraphics[width=\figWc,height=\figHc,angle=0]{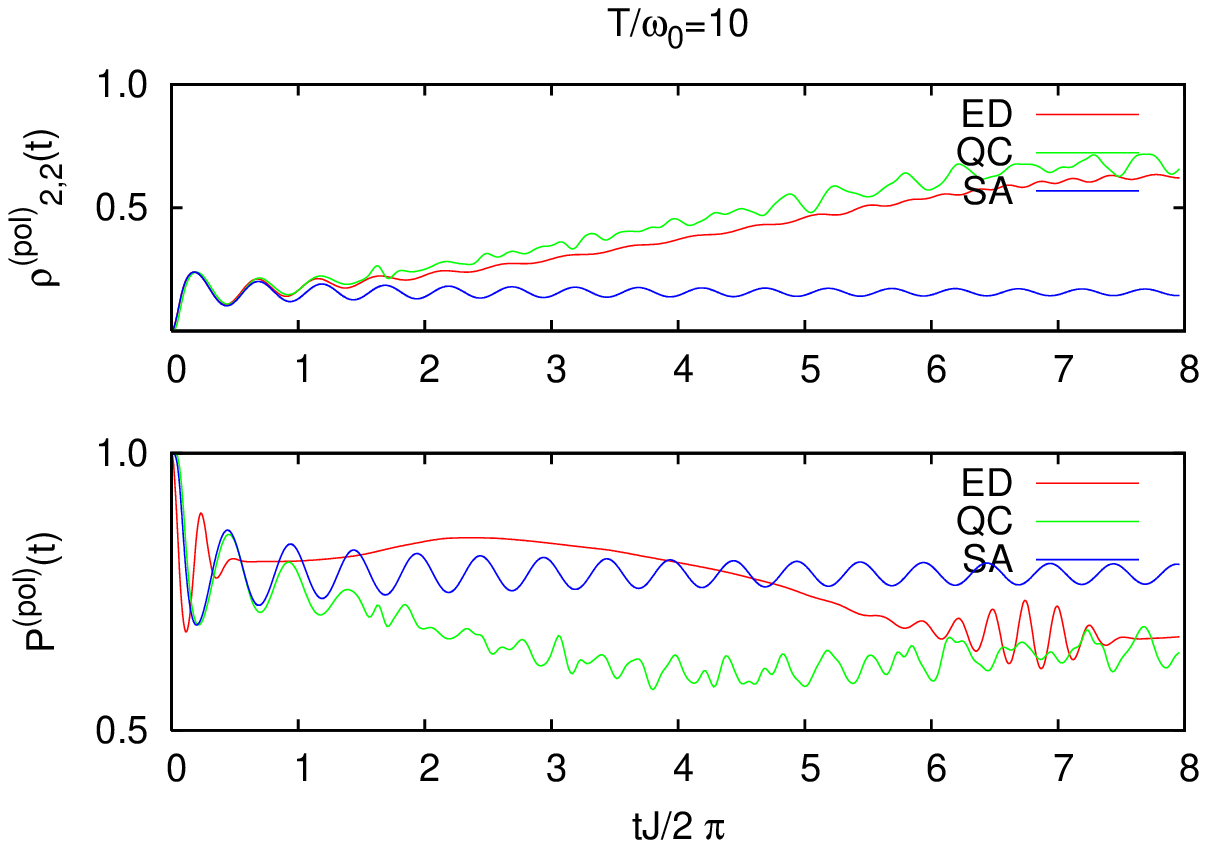}
\end{center}
\caption{Electron (left panels) and Polaron (right panels) dynamics at different temperatures
$T/\omega_0=0.1;1;10$. Adiabatic and strong coupling regime:$\gamma=0.1$ and $\lambda=2$.
For each temperature are shown the transition probability and the purity.
}
\label{fig:cfr_adiab_strong}
\end{figure}

\begin{figure}[htbp]
\begin{center}
\includegraphics[width=\figWc,height=\figHc,angle=0]{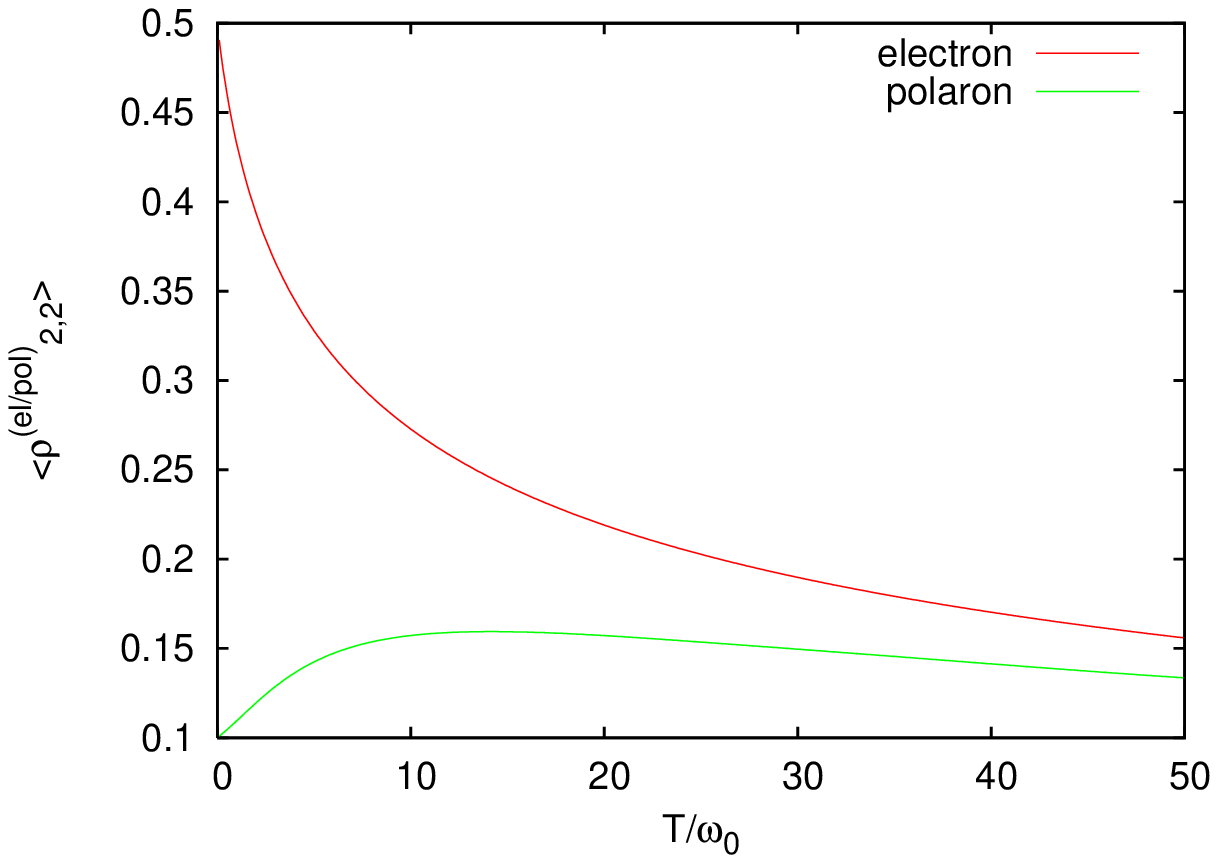}
\end{center}
\caption{Time average transition probability as function of time in the SA approximation.}
\label{fig:av}
\end{figure}
In this section, we show a comparison between the results obtained in the 
three different ways described before: exact diagonalization (ED) by means of 
mapping introduced in Sect. \ref{analytics}, the quantum-classical (QC) dynamics approach
described in Sect. \ref{app:QC} and the static Born-Oppenheimer (SA) approximation 
(Sect. \ref{sub:adiab}).
We shall limit ourselves to an adiabatic case ($\gamma=0.1$) 
with electron-phonon interaction strong enough to allow the polaron formation
($\lambda= 2$).

Let us start from the electron case (left panels in Fig. (\ref{fig:cfr_adiab_strong})).
At short times, we see that the state, not being an an eigenstate of $H$, starts to oscillate
coherently with frequency $J$, as expected for a not interacting electron.
Because of interaction, oscillations are damped and the electron starts to be entangled with phonon. 
The time-range considered here is of the order of $\omega_0^{-1}$. On this timescale
we can see that the higher  the temperature, the higher  the damping of the transition 
probability and, on the same time, the coherence goes down.
This means that the electronic state tends to localize on one of the two sites with an equal classical probability, in other words
it becomes a classical mixture of localized states.
This is an interesting example of  system going to a sort of equilibrium despite 
its finite size. This is because, in the adiabatic limit, the 
oscilltor spectrum tends to a continuum.
Actually, since the system is not exactly 
adiabatic, a partial recoherence is gained over much larger timescales of the order of the level spacing of the interactting system.
Analysis of this regime is beyond the scope of the present paper and will appear in a forthcoming publication.
Let us now compare the three different techniques. At low temperature the classical phonon is almost 
freezed, and so both SA and QC approximation are equivalent. Nevertheless, the ED
behaviour is quite different because in the exact dynamics also quantum fluctuations play a role.
A transition to a semiclassical behaviour can be seen for $T$ is comparable with $\omega_0$
when SA and  QC reproduce the ED dynamics.
Notice that the same occurs in the transport of extended system where classical incoherent transport 
is achieved when $T$ is greater than $0.2 \omega_0$ \cite{frat2} so that when $T\simeq\omega_0$ we have a classical incoherent transport.
 
At high temperature the oscillator dynamics plays a relevant role, QC is a much better approximation of ED than SA.
This fact can be understood by noting that the main temperature effect is
the damping of the coherent tunnelling oscillations.
Once these oscillation are sufficiently suppressed, the phonon driven dynamics prevails.
In the SA framework, the initial thermal distribution of the phonon coordinate 
makes the electron thermalizes irreversibly, in a time that is the shorter  the greater  the temperature.
Before this adiabatic thermalization, i.e, in a tunnelling period, the SA is still a good  approximation. Afterwards, the tunnelling dynamics saturates and the phonon dynamics has to be taken into account.

As far as the polaron is concerned, we can see (right panels in Fig.(\ref{fig:cfr_adiab_strong})), that, for
low temperature, the polaron  is still a good picture, keeping the state
quite pure even at long times.
Nevertheless, the polaron is very localized and its transition probability
is extremely low. The polaron is trapped inside the initial site.
As the temperature increases,  state becomes mixed but, in contrast with the 
electron, the transition probability increases.
So, in the adiabatic limit, the polaron mobility is increased by 
a thermal activation. 
That can be also seen in Fig. (\ref{fig:av}), where we reported the SA time averages
of the probability for the charge to be on site $2$, in both electron and polaron cases. As one can see, the curve is monotonic decreasing for the electron, while the polaron mobility starts to be enhanced  by the temperature before going down as for the electron.  

As it was stressed before, SA ad QC
do not provide a good limit for the purity, as we can verify 
by looking the right panels in Fig. (\ref{fig:cfr_adiab_strong})).
Nevertheless, for the population dynamics, are still valid the same considerations done for the electron.

\section{conclusion}
We introduced a density matrix approach for characterizing the
reduced charge dynamics in a two-site Holstein model.
Both the single site population and the purity are studied in order to
connect the charge transfer with its coherence.
In the adiabatic limit, two common approximation has been compared with exact 
results, the aim is to highlight the limits of validity of these approximations and
to provide a simple testing tool, the two-site model, for generalizations to other extended models.
Electron and polaron dynamics has been compared. Despite the two-level description of the moving 
charge,  temperature induces decoherence on 
intermediate timescale, due to the strong interaction with the oscillator. 
In the polaron case, temperature can enhance the (incoherent) charge transfer.

\bibliographystyle{iopart-num}
\bibliography{/home/simone/universita/bibliografia/bibliografia,/home/simone/universita/bibliografia/bibl-2site-coer,/home/simone/universita/bibliografia/bibl-electron-phon}
\end{document}